\documentclass[a4paper,fleqn,usenatbib]{mnras}
\usepackage[utf8]{inputenc}

\usepackage{newtxtext,newtxmath}
\usepackage{multirow}

\usepackage[T1]{fontenc}
\usepackage{ae,aecompl}

\usepackage{graphicx}	
\usepackage{amsmath}	
\usepackage{amssymb}	

\def\arcdeg{\hbox{$^\circ$}}

\def\arcsec{\hbox{$^{\prime\prime}$}}

\def\xte{XTE\,J1829-098}

\newcommand {\be}{\begin {equation}}
\newcommand {\ee}{\end {equation}}

\title[Discovery of CRSF in XTE\,J1829-098]{{\scshape Discovery of a
cyclotron absorption line in the transient X-ray pulsar XTE\,J1829-098}}

\author[A.\,E.\,Shtykovsky et al.]{
A.\,E.\,Shtykovsky$^{1}$\thanks{E-mail: a.shtykovsky@iki.rssi.ru},
A.\,A.\,Lutovinov$^{1}$,
S.\,S.\,Tsygankov$^{2,1}$
and S.\,V.\,Molkov$^{1}$
\\
$^{1}$Space Research Institute of the Russian Academy of Sciences, Profsoyuznaya Str. 84/32, Moscow 117997, Russia\\
$^{2}$Department of Physics and Astronomy, FI-20014 University of Turku, Finland}

\date{Accepted XXX. Received YYY; in original form ZZZ}

\pubyear{2018}

\begin{document}
\label{firstpage}
\pagerange{\pageref{firstpage}--\pageref{lastpage}}
\maketitle

\begin{abstract}
We report results of a spectral and timing analysis of the X-ray pulsar \xte\
using data obtained with the {\it NuSTAR} observatory during an outburst in
August 2018. A strong absorption feature was detected at the energy of
$E_{cyc}\simeq 15$ keV in the source spectrum. This feature was interpreted
as a cyclotron resonance scattering line corresponding to the magnetic field
strength of the neutron star surface $B\simeq1.7\times10^{12}$ G. The pulse
phase-resolved spectroscopy shows that the cyclotron line is significantly
detected at all phases of the pulse and its energy and other parameters are
variable over the pulse period. The timing analysis of the source emission
revealed strong pulsations with a period of $P = 7.84480(2)$ s. The pulsed
fraction is changed with the energy, including its local increase in the
vicinity of the cyclotron line. Using the archival data of the {\it RXTE}
observatory the presence of the cyclotron line in the spectrum of \xte\ was
independently confirmed.

\end{abstract}




\section{Introduction}

XTE\,J1829-098 was discovered by the \textit{RXTE} observatory during scans
of the Galactic plane in July 2004. It has been identified as a transient
X-ray pulsar with the pulse period of $\sim7.8$ s
\citep{2004ATel..317....1M}.

Based on the \textsl{XMM-Newton} and \textsl{Chandra} data
\citet{2007ApJ...669..579H} showed that the spectrum of XTE\,J1829-098 in
soft X-rays ($<10$ keV) can be described by an absorbed powerlaw model. It is
interesting to note, that \textsl{XMM-Newton} detected XTE\,J1829-098
serendipitously on March, 2003 (i.e. before its formal discovery) during the
Galactic Plane Survey program, whereas \textsl{Chandra} observations were
performed three times in 2007. Data of the \textsl{Chandra} observatory
allowed \citet{2007ApJ...669..579H} to localize the source with an accuracy
of 0.6\arcsec\ and to determine its infrared counterpart, but the type and
class of this star are still unclear. These authors claimed also that the
observed X-ray luminosity in the 2-10~keV energy band $L_{\rm X} \simeq 2
\times 10^{36} (d/10 \text{ kpc})^2 \text{ erg s}^{-1}$, where $d$ is the
distance to the source, is typical for Be X-ray transients or wind-fed
systems.

The flux registered from XTE\,J1829-098 is varied over the range of more than
three orders of magnitude, that points to the transient nature of the source.
Using a long term light curve of the source from {\it RXTE}/PCA monitoring
observations, \citet{2009ATel.2007....1M} estimated an expected recurrence
period of $\sim246$ days and an outbursts duration of $\sim7$ days.

New outburst from XTE\,J1829-098 was detected with the {\it MAXI} monitor on
Aug 5, 2018 \citep{2018ATel11927....1N} with the flux of $\simeq24$ mCrab in
the 4-10 keV energy band. Immediately after the {\it MAXI} detection a TOO
observation with the {\it NuSTAR} observatory was triggered with the main
purpose to reconstruct the broad band energy spectrum of the source and to
search for the cyclotron line.

In Section\,\ref{obs} we describe observations, which were used in the paper,
and the data reduction procedure. Results of the timing analysis, including
pulse period measurements and pulse profile studies, are presented in
Section\,\ref{timing}. In Section\,\ref{sec:spec} we report a discovery of
the cyclotron absorption line in the spectrum of XTE\,J1829-098, results of
the pulse phase-resolved spectroscopy and an independent confirmation of the
cyclotron line detection from {\it RXTE} data. Results are summarized and
briefly discussed in Section\,\ref{concl}.

\section{Observations and data reduction}
\label{obs}

Observations of XTE\,J1829-098 were performed with \textsl{NuSTAR} on Aug 16,
2018 (ObsID 90401332002) with an on-source exposure time of $\sim 27.8$ ks
and an average count rate of $\sim8$\,cts s$^{-1}$ per module.

The \textsl{NuSTAR} observatory consists of two identical X-ray telescope
modules, each equipped with independent mirror systems and focal plane
detector units, also referred to as FPMA and FPMB
\citep{2013ApJ...770..103H}. It provides X-ray imaging, spectroscopy and timing in
the energy range of 3--79~keV with an angular resolution of 18\arcsec~(FWHM)
and spectral resolution of 400~eV (FWHM) at 10~keV.

\begin{figure}
\includegraphics[width=0.85\columnwidth,bb=82 82 455 597,clip]{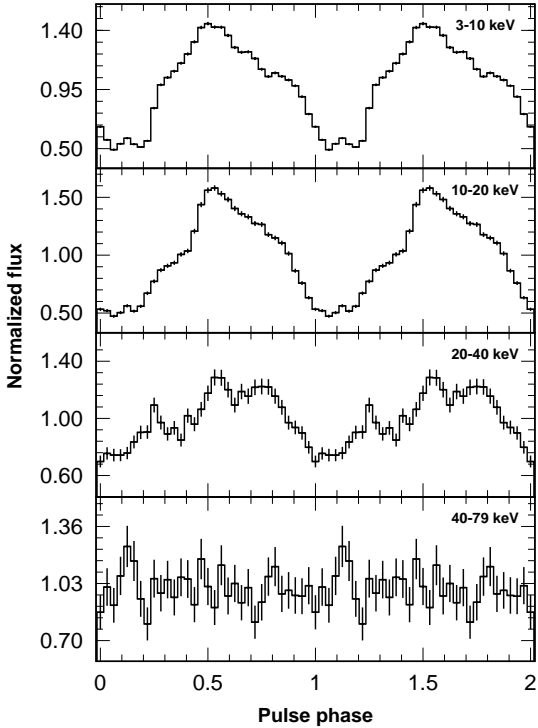}
\caption{Pulse profile of \xte\ in 3-10~keV, 10-20~keV,
20-40~keV and 40-79~keV energy bands.} \label{fig:ppbins}
\end{figure}

To extract spectra and light curves we used the standard \textsc{nustardas}
1.8.0 software as distributed with the \textsc{HEASOFT} 6.24 package and the
\textsc{CALDB} version 20180814. The standard \textsc{lcmath} tool was used
to combine the light curves of the \textsl{NuSTAR} modules to improve a
statistic for timing analysis. The source data were extracted from a circular
region with radius of $130$\arcsec, centered at the source position. The
background data were extracted using a polygonal region away from the source
position. It is important to note that the observational data show no signs
of a contamination by a stray-light or ghost rays.

Additionally data of the {\it RXTE} observatory \citep{1993A&AS...97..355B}
were used for an independent analysis and confirmation of the {\it NuSTAR}
detections. The source was observed with {\it RXTE} in the pointing mode
about two dozen times in three epochs: Aug 2004 (ObsID 90058), Aug 2008
(ObsID 93445) and Apr 2009 (ObsID 94419). We used here only data of the
Proportional Counter Array \citep[PCA,][]{2006ApJS..163..401J} stored in the
{\sc standard2} mode.

The obtained spectra were grouped to have more than 20 counts per bin using
the \texttt{grppha} tool. The final data analysis (timing and spectral) was
performed with the {\sc HEASOFT 6.24} software package. All uncertainties are
quoted at the $1\sigma$ confidence level, if otherwise stated.

\section{Spin period and pulse profile}
\label{timing}

\begin{figure}
\includegraphics[width=0.95\columnwidth,bb=33 272 553 678]{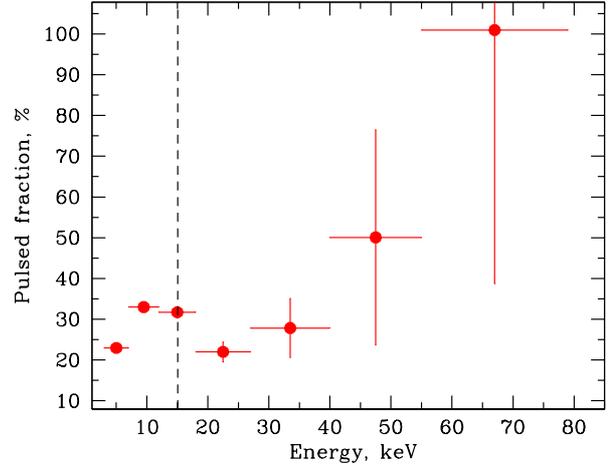}
\caption{Dependence of the pulsed fraction of \xte\ on the energy. The dashed
line shows the position of the cyclotron line found in the source
spectrum.}\label{fig:pf}
\end{figure}

The pulse period of the source was determined by using the epoch-folding
technique (the \texttt{efsearch} tool of the {\sc HEASOFT} package) we
measured the pulse period as $P_{\rm spin} = 7.84480 \pm 0.00002 \text{ s}$.
The uncertainty was defined by the bootstrapping technique \citep[see][for
details]{2013AstL...39..375B}.

Source light curves in different energy bands were folded into 32 phase bins
using the best-fit spin period value. Corresponding pulse profiles in
3--10~keV, 10--20~keV, 20--40~keV and 40--79~keV energy bands are shown in
Fig.\,\ref{fig:ppbins}. Generally, the pulse profile has a quite smooth and
simple shape similar to an asymmetric sin-like one. In the 40--79~keV energy
band the photon statistics is quite poor and the pulse profile is dominated
mostly by a noise.

The dependence of the pulsed fraction\footnote{The pulsed fraction is defined
as $\textrm{PF} = (I_{\rm max} - I_{\rm min}) / (I_{\rm max} + I_{\rm min})$,
where $I_{\rm max}$ and $I_{\rm min}$ are maximum and minimum intensities in
the pulse profile, respectively.} on the energy is presented in
Fig.\,\ref{fig:pf}. At low energies its value is about 20\% with a gradual
increase at higher energies, that is typical for the majority of X-ray
pulsars \citep{2009AstL...35..433L}. Besides this general tendency the
dependence shows another prominent feature: a local increase of the pulsed
fraction up to 30-35\% in the energy range of 10-18 keV. Such increases were
initially revealed in two brightest transient pulsars V\,0332+53 and
4U\,0115+63 in the vicinity of the cyclotron line energies and their
harmonics \citep{2006MNRAS.371...19T,2007AstL...33..368T}. Later similar
features were also found for a number of other X-ray pulsars
\citep{2009AstL...35..433L}. Thus, the peculiar increase of the pulsed
fraction can be considered as an indication of a possible presence of the
cyclotron line at energies 10-18 keV.

\section{Spectral analysis}
\label{sec:spec}

\subsection{Phase-averaged spectroscopy with {\it NuSTAR}}

\begin{figure}
\centering
\includegraphics[width=0.88\columnwidth,bb=60 55 598 599,clip]{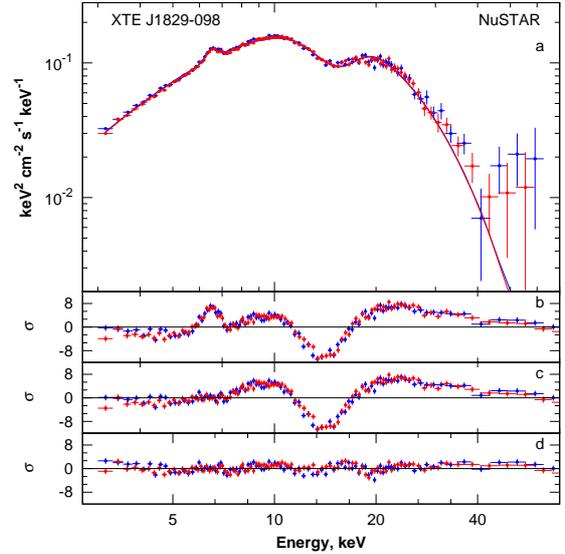}
\caption{Energy spectrum of XTE\,J1829-098 measured with \textsl{NuSTAR} (a):
red color corresponds to the FPMA data, blue one to the FPMB data. Residuals
of the models: \texttt{cutoffpl} (b), \texttt{cutoffpl + gauss} (c) and
\texttt{cutoffpl*gabs+gauss} (d).}\label{fig:spec}
\end{figure}

The spectrum of \xte\ is typical for accreting X-ray pulsars and demonstrates
a cutoff at high energies (Fig.\,\ref{fig:spec}a). At the first stage it was
approximated with several commonly used models: a power law with an
exponential cutoff in the form $E^{-\Gamma}{\rm exp}(-E/E_{\rm fold})$ (\texttt{cutoffpl} in the
{\sc XSPEC} package), a power law with a high energy cutoff
(\texttt{powerlaw*highcut} in the {\sc XSPEC} package) and a thermal
Comptonization (\texttt{comptt} in the {\sc XSPEC} package). The source and
background spectra from both modules of {\it NuSTAR} were used for
simultaneous fitting. To take into account the uncertainty in the instrument
calibrations cross-calibration constant $C$ between them was included in all
spectral models.

None of the above mentioned models describe spectrum well: there are positive
deviations of the data at energies 6-7 keV and strong negative deviations at
energies 13-18 keV (Fig.\,\ref{fig:spec}b). In the following analysis we used
the \texttt{cutoffpl} model as it describes the spectrum better and has fewer
parameters. An addition to the model an iron line at $\sim6.4$ keV in the
gaussian form improves the fit, but its quality is still non-acceptable
($\chi^2=4770$ for 1418 d.o.f.) due to a deficit of photons around 15 keV
(Fig.\,\ref{fig:spec}c). To describe this feature an absorption component in
the form of the \texttt{gabs} model was added to the model. It
led to a significant improvement of the fit quality ($\chi^2=1551$ for 1415
d.o.f., Fig.\,\ref{fig:spec}d) and adequate description of the source
spectrum.

\begin{table}
\caption{Best-fit parameters of the XTE\,J1829-098 spectrum according to
the \textsl{NuSTAR} data}\label{tab:sparam}


	\centering
	\footnotesize

	\begin{tabular}{lc}
		\hline\hline
		Parameter & Value \\
		\hline
		 $\Gamma$ 			       & $-0.76 \pm 0.02$ \\
		 $E_{\rm fold}$, keV      & $ 4.45 \pm 0.03$ \\
		 $E_{\rm cyc}$, keV      & $15.05 \pm 0.06$ \\
		 $W_{\rm cyc}$, keV      & $ 2.26 \pm 0.06$ \\
		 $\tau_{\rm cyc}$   & $ 0.52 \pm 0.05$ \\
		 $E_{\rm Fe}$, keV        & $ 6.50 \pm 0.02$ \\
		 $\sigma_{\rm Fe}$, keV   & $ 0.24 \pm 0.02$ \\
         $N_{\rm Fe}$               & $ (2.96 \pm 0.19)\times10^{-4}$ \\
         $C$                       & $ 1.014 \pm 0.003$ \\
		\hline
	\end{tabular}

\end{table}

Thus, the spectrum of \xte\ can be well approximated by the \texttt{cutoffpl}
model modified by the emission line, associated with the fluorescent iron
line, and the absorption line with the energy of $\simeq15$ keV. Best-fit
parameters of this model are summarized in Table\,\ref{tab:sparam}.

\begin{figure}
\centering
\includegraphics[width=0.9\columnwidth,bb=19 235 497 711,clip]{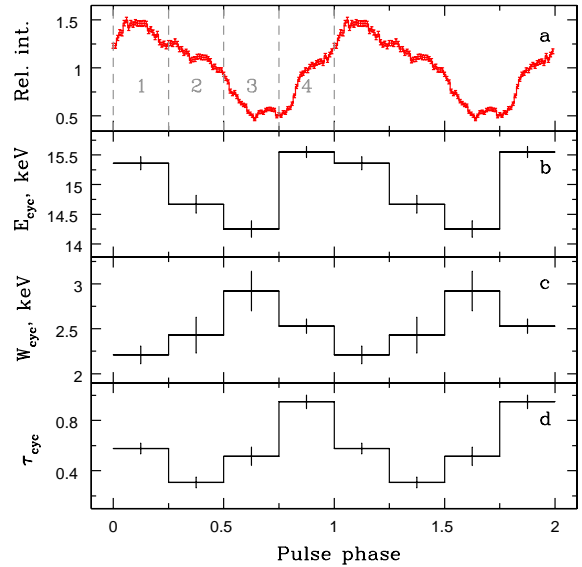}
\caption{Pulse profile of \xte\ in the 3-79 energy band (a).
Variations of the cyclotron line parameters (b, c, d) as a
function of the pulse phase (a).} \label{fig:phspec}
\end{figure}

The absorption feature at 15 keV can be interpreted as a possible cyclotron
resonant scattering feature. To approximate cyclotron absorption lines two
models from the {\sc XSPEC} package are usually used: \texttt{gabs} and
\texttt{cyclabs}. Both models describe the data adequately well, but the
cyclotron line energy derived from the \texttt{cyclabs} model is
systematically lower  than the energy derived from the \texttt{gabs} one
\citep[see, e.g.,][]{2012MNRAS.421.2407T, 2015MNRAS.448.2175L}. If one would
use the \texttt{cyclabs} model for the description of the \xte\ spectrum than
the cyclotron line energy would be $E_{\rm cyc}\simeq14.2$ keV.

In spectra of several X-ray pulsars higher harmonics of the cyclotron line
are also registered. To examine a presence of higher harmonics in the
spectrum of \xte\ we added to the model an absorption component at the energy
corresponding to the double cyclotron line energy; the line width was fixed
at the value from Table\,\ref{tab:sparam}. No further improvement of the fit
was found and only an upper limit for the optical depth of the first harmonic
of the cyclotron line was obtained as 0.05 (90\% confidence level).

An averaged X-ray flux in the 3--79~keV energy band during the {\it NuSTAR}
observation was $F_{\rm x} \simeq 3.6 \times 10^{-10} \text{ erg cm}^{-2}
\text{ s}^{-1}$, that corresponds to the source luminosity of $L_{\rm x}
\simeq 4.3 \times 10^{36} (d/10 \text{ kpc})^2 \text{ erg s}^{-1}$. Taking
into account the spectrum shape it corresponds approximately to the flux
measured by \citet{2007ApJ...669..579H} in the 2-10 keV energy band.

\subsection{Pulse phase-resolved spectroscopy}

\begin{figure}
\centering
\includegraphics[width=0.95\columnwidth,bb=40 180 543 686,clip]{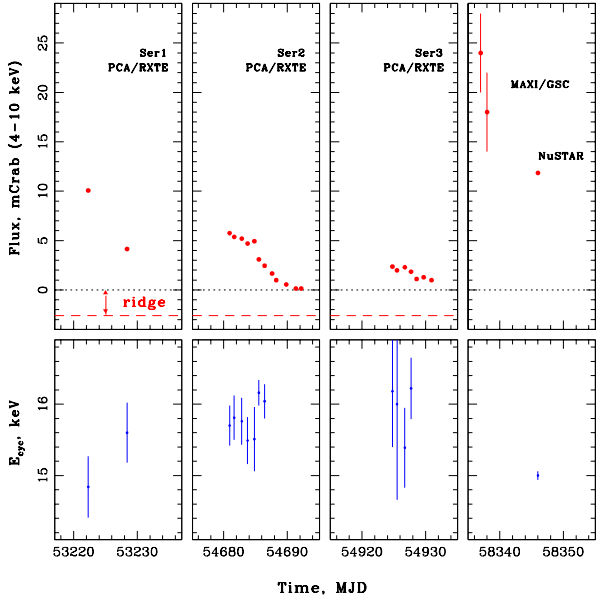}
\caption{Upper panels: Evolution of the \xte\ flux during four outbursts.
Bottom panels: Measured cyclotron line energy.}\label{fig:line_evol}
\end{figure}

In order to study the evolution of spectral parameters of \xte\ on the time
scale of the pulse period a pulse phase-resolved spectroscopy
was carried out. The pulse period was divided into four intervals and
corresponding spectra were extracted from the data. To approximate these
spectra we applied the same model that was used for the phase-averaged
spectroscopy. As for the phase-averaged spectrum the cyclotron absorption
line is detected in all phase bins with high significance.

An evolution of the line parameters with the pulse phase is shown in
Fig.\,\ref{fig:phspec}. The cyclotron line energy is significantly (more than
1 keV) varied over the pulse and approximately correlates with the pulse
intensity, reaching values 15.3-15.5 keV at the pulse rise and peak and
decreasing down to $\sim14.3$ keV at the pulse minimum. The cyclotron line
width is varied nearly sinusoidally in opposition to the pulse intensity,
whereas the line optical depth peaks at the maximum of the line energy. Such
variations of line parameters over the pulse are quite usual, observed in
many X-ray pulsars and probably connected with changes in the viewing angle
to the regions where the cyclotron line is formed \citep[see,
e.g.,][]{2015MNRAS.448.2175L}.

\subsection{Spectroscopy of \xte\ with {\it RXTE}}

To examine independently results obtained with the {\it NuSTAR} observatory
and to investigate a possible evolution of the cyclotron line energy
with time and the source luminosity we used data from the {\it RXTE}
observatory, which observed \xte\ many times in 2004-2009.

The source flux history in the 4-10 keV energy band measured during four
outbursts (three ones were observed with {\it RXTE}/PCA and one was observed
with {\it MAXI} and {\it NuSTAR}) is shown on upper panels of
Fig.\,\ref{fig:line_evol}. Fluxes, measured by {\it MAXI} and {\it NuSTAR},
were taken from \citet{2018ATel11927....1N} and the current work,
respectively. A correct extraction of the source spectra from the {\it RXTE} data
and calculation of corresponding fluxes meet some peculiarities due to the
source location close to the Galactic plane, that are described in details in
the Appendix.

\begin{figure}
\centering
\includegraphics[width=0.98\columnwidth,bb=61 181 495 623,clip]{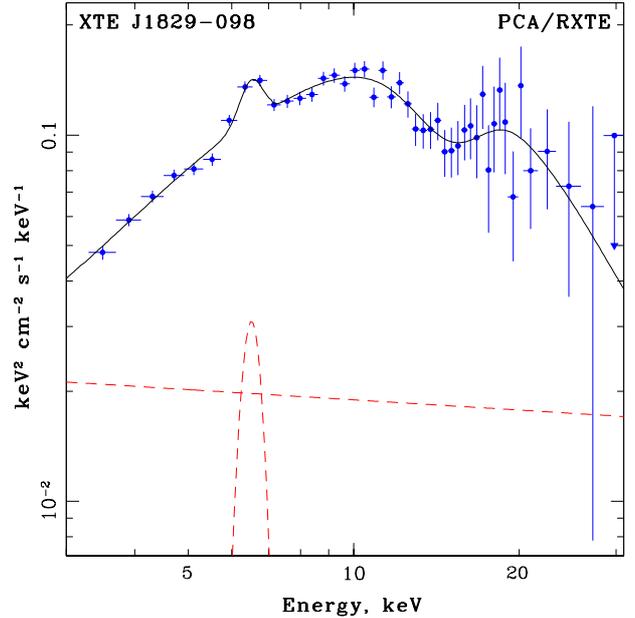}
\caption{The energy spectrum of \xte\ measured by the {\it RXTE}/PCA
instrument in the brightest state of the source (the first point in
Fig.~\ref{fig:line_evol}). Red dashed lines show a contribution of the
Galactic ridge component to the total PCA spectrum (see the text for
details). Black solid line represents the best fit model for the total
spectrum measured with PCA.} \label{fig:pca_spc}
\end{figure}

To approximate the \xte\ spectra obtained with {\it RXTE}/PCA we used the
same spectral model as for the {\it NuSTAR} observation. Our analysis
revealed that all spectra are well approximated with this model and an
inclusion of the absorption component at energies 15-16 keV is required to
improve the fit quality in comparison with the simple \texttt{cutoffpl}
model. Thus we can independently confirm the presence of the cyclotron
absorption line in the spectrum of \xte. Due to the source faintness and
short exposures of {\it RXTE} observations all three parameters of the
cyclotron line cannot be well restricted. Therefore we fixed the line width
$W_{\rm cyc}$ on the value determined by {\it NuSTAR} for the averaged
spectrum. Results of measurements of the cyclotron line energy with the {\it
RXTE} observatory are presented on bottom panels of
Fig.\,\ref{fig:line_evol}. The figure demonstrates that the cyclotron line
energy is in good agreement for the first {\it RXTE}/PCA observation and
{\it NuSTAR} one, when the source flux has a comparable level, and
systematically higher (but statistically insignificant) for fainter
outburst phases.

\section{Conclusions}
\label{concl}

In this Letter we report a discovery of the cyclotron absorption line at
$E_{\rm cyc} \simeq 15$ keV in the spectrum of the transient X-ray pulsar
\xte. Such features are registered in spectra of many X-ray pulsars
\citep[see, e.g., recent review of][]{2015A&ARv..23....2W} and usually used
for the direct estimates of the magnetic fields of neutron stars. The
measured energy of the cyclotron absorption line corresponds to the magnetic
field strength $B \simeq 1.7 \times 10^{12}$~G on the surface of the neutron
star, that is typical for X-ray pulsars.

The pulse phase-resolved spectroscopy revealed that parameters of the
cyclotron line are variable over the pulse, that probably connected with
changes of the viewing angle to the regions where the cyclotron line is
formed.

The study of the pulse profile and pulsed fraction dependencies on the energy
revealed two distinct features: an overall increase of the pulsed fraction
with the energy and its local enhancement at energies 10-18 keV, i.e. near
the cyclotron line energy. Similar peculiarities of the pulsed fraction were
observed earlier for several other X-ray pulsars.

Finally, using the archival data of the {\it RXTE} observatory the presence
of the cyclotron line in the spectrum of \xte\ was independently confirmed.
Moreover some hint to possible variations of the line energy with the flux
was revealed, but for the final conclusions special monitoring observations
with the high sensitivity instruments (like the {\it NuSTAR} one) are required.

\section*{Acknowledgements}

This work was supported by the grant of the Ministry of Science and High
Education 14.W03.31.0021. The research has made by using data obtained with
\textsl{NuSTAR}, a project led by Caltech, funded by NASA and managed by
NASA/JPL, and has utilized the \textsc{nustardas} software package, jointly
developed by ASDC and Caltech. AS would like to thank Karl
Forster for the assistance in scheduling the observation.




\bibliographystyle{mnras}
\bibliography{xte_j1829m098}

\begin{thebibliography}{}
\makeatletter
\relax
\def\mn@urlcharsother{\let\do\@makeother \do\$\do\&\do\#\do\^\do\_\do\%\do\~}
\def\mn@doi{\begingroup\mn@urlcharsother \@ifnextchar [ {\mn@doi@}
  {\mn@doi@[]}}
\def\mn@doi@[#1]#2{\def\@tempa{#1}\ifx\@tempa\@empty \href
  {http://dx.doi.org/#2} {doi:#2}\else \href {http://dx.doi.org/#2} {#1}\fi
  \endgroup}
\def\mn@eprint#1#2{\mn@eprint@#1:#2::\@nil}
\def\mn@eprint@arXiv#1{\href {http://arxiv.org/abs/#1} {{\tt arXiv:#1}}}
\def\mn@eprint@dblp#1{\href {http://dblp.uni-trier.de/rec/bibtex/#1.xml}
  {dblp:#1}}
\def\mn@eprint@#1:#2:#3:#4\@nil{\def\@tempa {#1}\def\@tempb {#2}\def\@tempc
  {#3}\ifx \@tempc \@empty \let \@tempc \@tempb \let \@tempb \@tempa \fi \ifx
  \@tempb \@empty \def\@tempb {arXiv}\fi \@ifundefined
  {mn@eprint@\@tempb}{\@tempb:\@tempc}{\expandafter \expandafter \csname
  mn@eprint@\@tempb\endcsname \expandafter{\@tempc}}}

\bibitem[\protect\citeauthoryear{{Boldin}, {Tsygankov}  \&
  {Lutovinov}}{{Boldin} et~al.}{2013}]{2013AstL...39..375B}
{Boldin} P.~A.,  {Tsygankov} S.~S.,   {Lutovinov} A.~A.,  2013, \mn@doi
  [Astronomy Letters] {10.1134/S1063773713060029}, \href
  {http://adsabs.harvard.edu/abs/2013AstL...39..375B} {39, 375}

\bibitem[\protect\citeauthoryear{{Bradt}, {Rothschild}  \& {Swank}}{{Bradt}
  et~al.}{1993}]{1993A&AS...97..355B}
{Bradt} H.~V.,  {Rothschild} R.~E.,   {Swank} J.~H.,  1993, \aaps, \href
  {http://adsabs.harvard.edu/abs/1993A%26AS...97..355B} {97, 355}

\bibitem[\protect\citeauthoryear{{Halpern} \& {Gotthelf}}{{Halpern} \&
  {Gotthelf}}{2007}]{2007ApJ...669..579H}
{Halpern} J.~P.,  {Gotthelf} E.~V.,  2007, \mn@doi [\apj] {10.1086/521704},
  \href {http://adsabs.harvard.edu/abs/2007ApJ...669..579H} {669, 579}

\bibitem[\protect\citeauthoryear{{Harrison} et~al.,}{{Harrison}
  et~al.}{2013}]{2013ApJ...770..103H}
{Harrison} F.~A.,  et~al., 2013, \mn@doi [\apj] {10.1088/0004-637X/770/2/103},
  \href {http://adsabs.harvard.edu/abs/2013ApJ...770..103H} {770, 103}

\bibitem[\protect\citeauthoryear{{Jahoda}, {Markwardt}, {Radeva}, {Rots},
  {Stark}, {Swank}, {Strohmayer}  \& {Zhang}}{{Jahoda}
  et~al.}{2006}]{2006ApJS..163..401J}
{Jahoda} K.,  {Markwardt} C.~B.,  {Radeva} Y.,  {Rots} A.~H.,  {Stark} M.~J.,
  {Swank} J.~H.,  {Strohmayer} T.~E.,   {Zhang} W.,  2006, \mn@doi [\apjs]
  {10.1086/500659}, \href {http://adsabs.harvard.edu/abs/2006ApJS..163..401J}
  {163, 401}

\bibitem[\protect\citeauthoryear{{Lutovinov} \& {Tsygankov}}{{Lutovinov} \&
  {Tsygankov}}{2009}]{2009AstL...35..433L}
{Lutovinov} A.~A.,  {Tsygankov} S.~S.,  2009, \mn@doi [Astronomy Letters]
  {10.1134/S1063773709070019}, \href
  {http://adsabs.harvard.edu/abs/2009AstL...35..433L} {35, 433}

\bibitem[\protect\citeauthoryear{{Lutovinov}, {Tsygankov}, {Suleimanov},
  {Mushtukov}, {Doroshenko}, {Nagirner}  \& {Poutanen}}{{Lutovinov}
  et~al.}{2015}]{2015MNRAS.448.2175L}
{Lutovinov} A.~A.,  {Tsygankov} S.~S.,  {Suleimanov} V.~F.,  {Mushtukov} A.~A.,
   {Doroshenko} V.,  {Nagirner} D.~I.,   {Poutanen} J.,  2015, \mn@doi [\mnras]
  {10.1093/mnras/stv125}, \href
  {http://adsabs.harvard.edu/abs/2015MNRAS.448.2175L} {448, 2175}

\bibitem[\protect\citeauthoryear{{Markwardt}, {Swank}  \& {Smith}}{{Markwardt}
  et~al.}{2004}]{2004ATel..317....1M}
{Markwardt} C.~B.,  {Swank} J.~H.,   {Smith} E.~A.,  2004, The Astronomer's
  Telegram, \href {http://adsabs.harvard.edu/abs/2004ATel..317....1M} {317}

\bibitem[\protect\citeauthoryear{{Markwardt}, {Halpern}  \&
  {Swank}}{{Markwardt} et~al.}{2009}]{2009ATel.2007....1M}
{Markwardt} C.~B.,  {Halpern} J.,   {Swank} J.~H.,  2009, The Astronomer's
  Telegram, \href {http://adsabs.harvard.edu/abs/2009ATel.2007....1M} {2007}

\bibitem[\protect\citeauthoryear{{Nakajima} et~al.,}{{Nakajima}
  et~al.}{2018}]{2018ATel11927....1N}
{Nakajima} M.,  et~al., 2018, The Astronomer's Telegram, \href
  {http://adsabs.harvard.edu/abs/2018ATel11927....1N} {11927}

\bibitem[\protect\citeauthoryear{{Revnivtsev}, {Sazonov}, {Gilfanov},
  {Churazov}  \& {Sunyaev}}{{Revnivtsev} et~al.}{2006}]{2006A&A...452..169R}
{Revnivtsev} M.,  {Sazonov} S.,  {Gilfanov} M.,  {Churazov} E.,   {Sunyaev} R.,
   2006, \mn@doi [\aap] {10.1051/0004-6361:20054268}, \href
  {http://adsabs.harvard.edu/abs/2006A%26A...452..169R} {452, 169}

\bibitem[\protect\citeauthoryear{{Tsygankov}, {Lutovinov}, {Churazov}  \&
  {Sunyaev}}{{Tsygankov} et~al.}{2006}]{2006MNRAS.371...19T}
{Tsygankov} S.~S.,  {Lutovinov} A.~A.,  {Churazov} E.~M.,   {Sunyaev} R.~A.,
  2006, \mn@doi [\mnras] {10.1111/j.1365-2966.2006.10610.x}, \href
  {http://adsabs.harvard.edu/abs/2006MNRAS.371...19T} {371, 19}

\bibitem[\protect\citeauthoryear{{Tsygankov}, {Lutovinov}, {Churazov}  \&
  {Sunyaev}}{{Tsygankov} et~al.}{2007}]{2007AstL...33..368T}
{Tsygankov} S.~S.,  {Lutovinov} A.~A.,  {Churazov} E.~M.,   {Sunyaev} R.~A.,
  2007, \mn@doi [Astronomy Letters] {10.1134/S1063773707060023}, \href
  {http://adsabs.harvard.edu/abs/2007AstL...33..368T} {33, 368}

\bibitem[\protect\citeauthoryear{{Tsygankov}, {Krivonos}  \&
  {Lutovinov}}{{Tsygankov} et~al.}{2012}]{2012MNRAS.421.2407T}
{Tsygankov} S.~S.,  {Krivonos} R.~A.,   {Lutovinov} A.~A.,  2012, \mn@doi
  [\mnras] {10.1111/j.1365-2966.2012.20475.x}, \href
  {http://adsabs.harvard.edu/abs/2012MNRAS.421.2407T} {421, 2407}

\bibitem[\protect\citeauthoryear{{Walter}, {Lutovinov}, {Bozzo}  \&
  {Tsygankov}}{{Walter} et~al.}{2015}]{2015A&ARv..23....2W}
{Walter} R.,  {Lutovinov} A.~A.,  {Bozzo} E.,   {Tsygankov} S.~S.,  2015,
  \mn@doi [\aapr] {10.1007/s00159-015-0082-6}, \href
  {http://adsabs.harvard.edu/abs/2015A%26ARv..23....2W} {23, 2}

\makeatother
\end{thebibliography}

\appendix

\section{\xte\ spectrum reconstruction with {\it RXTE}/PCA}

\begin{figure}
\centering
\includegraphics[width=0.87\columnwidth]{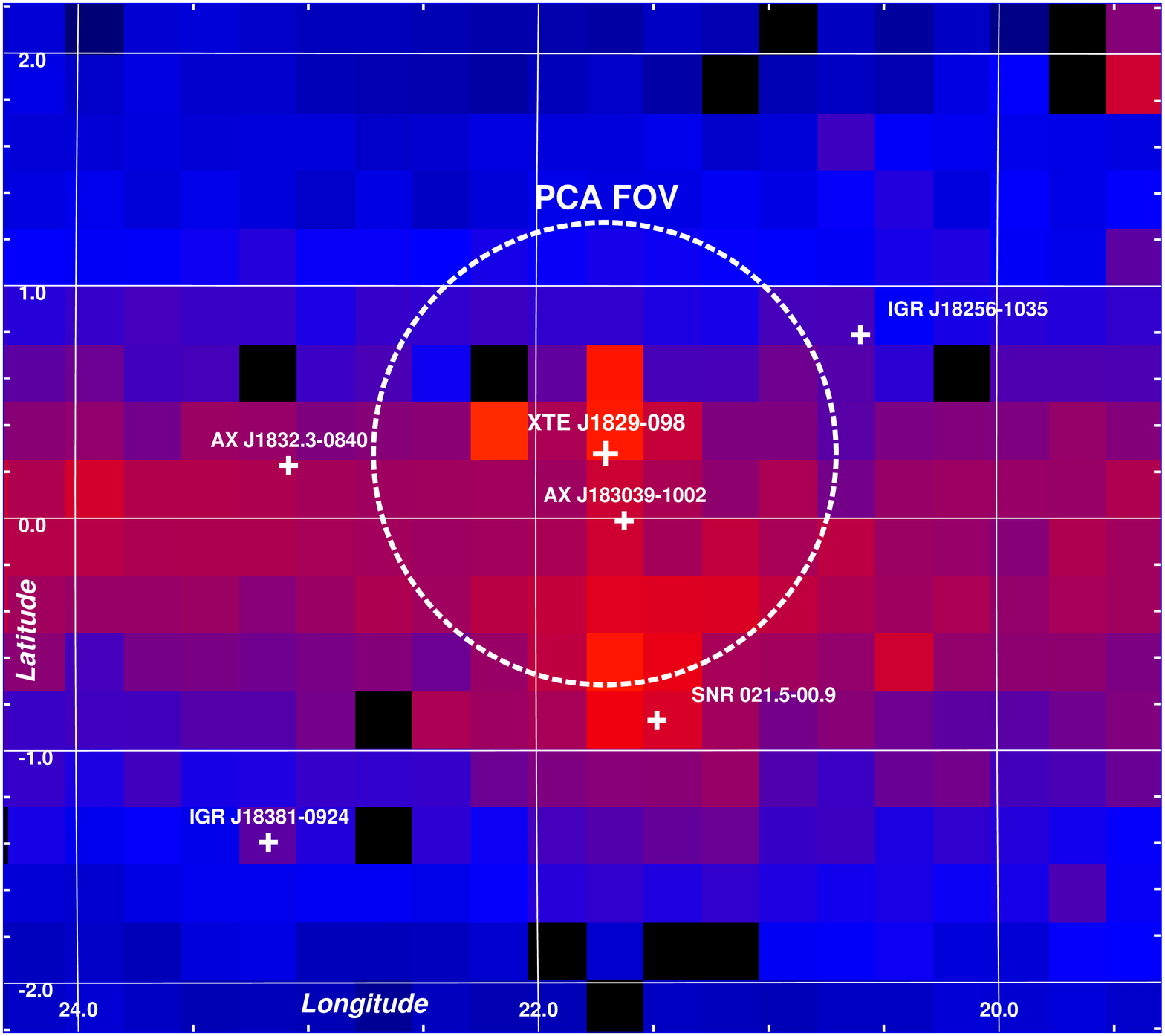}
\caption{The map of the sky around \xte, reconstructed with using
all Galactic scan observations. The dashed-line circle (1\arcdeg\ in radius)
shows the FOV of PCA during pointing observations of \xte.}
\label{fig:pca_fov}
\end{figure}

\xte\ is located in the sky area where the Galactic ridge emission could not
be treated as a negligible one for instruments with wide fields of view
similar to {\it RXTE}/PCA. The map of the sky region around \xte\ in the
total PCA energy band is shown in Fig.~\ref{fig:pca_fov}. This image had been
reconstructed using all PCA scanning observations of the Galactic plane. An
extended emission along the Galactic plane is clearly seen. Thus, to
reconstruct correctly the spectrum of \xte\ we should take into account not
only the instrumental background, but also a component connected with the
Galactic ridge emission and the possible contribution of individual sources
falling into the PCA FOV \citep[see, e.g.,][]{2004ATel..317....1M}. To
estimate this 'sky' background component and its spectrum we used two last
observations in the Ser. 2 (see upper panels of Fig.\,\ref{fig:line_evol}),
where \xte\ was on the undetectable level (i.e. pulsations from the source
were not detected). This 'sky' background emission can be adequately fitted
by the simple power law model with the spectral index of $2.1$, the Gaussian
emission line around $\sim 6.5$ keV and the total flux of $\simeq2.5$ mCrab
in the 4-10 keV energy band. This is well agreed with results of measurements
of the Galactic ridge emission by \citet{2006A&A...452..169R}. The spectrum
of the 'sky' background emission and its contribution to the total flux,
measured from the \xte\ with {\it RXTE}/PCA, are shown by red dashed lines in
Fig.\,\ref{fig:line_evol} and \ref{fig:pca_spc}.




%
%


\bsp	
\label{lastpage}
\end{document}